\documentclass[useAMS,usenatbib]{mn2e}

\usepackage{tabularx}
\usepackage{graphicx}
\usepackage{txfonts}
\usepackage{longtable}
\usepackage{hhline}
\usepackage{arydshln}
\usepackage{multirow}
\usepackage{lscape}
\usepackage{array}
\usepackage{rotating}
\usepackage{mathrsfs}

\newcommand{\hersc}{{\it Herschel}}

\newcommand{\lsun}{$L_\odot$}

\newcolumntype{R}[1]{>{\raggedleft\arraybackslash }b{#1}}
\newcolumntype{L}[1]{>{\raggedright\arraybackslash }b{#1}}
\newcolumntype{C}[1]{>{\centering\arraybackslash }b{#1}}

\newlength{\pointwidth}
\begin{document}
  
\title[SEPIA Band 5 line survey in Arp~220]{Water, methanol and dense gas tracers in the local ULIRG Arp~220: 
Results from the new SEPIA Band 5 Science Verification campaign}

\author[Galametz et al.]
{\parbox{\textwidth}{M. Galametz$^{1}$, 
Z.-Y. Zhang$^{2,1}$,
K. Immer$^{1}$,
E. Humphreys$^{1}$, 
R. Aladro$^{3,4}$,
C. De Breuck$^{1}$,
A. Ginsburg$^{1}$,
S. C. Madden$^{5}$,
P. M{\o}ller$^{1}$,
V. Arumugam$^{1}$
}\vspace{0.5cm}\\
\parbox{\textwidth}{
$^{1}$ European Southern Observatory, Karl-Schwarzschild-Str. 2, D-85748 Garching-bei-M\"unchen, Germany. Contact: maud.galametz@eso.org \\
$^{2}$ Institute for Astronomy, University of Edinburgh, Royal Observatory, Blackford Hill, Edinburgh EH9 3HJ, UK \\
$^{3}$ European Southern Observatory, Avda. Alonso de C\'{o}rdova 3107, Vitacura, Santiago, Chile \\
$^{4}$ Department of Earth and Space Sciences, Chalmers University of Technology, Onsala Observatory, 439 92 Onsala, Sweden \\
$^{5}$ Laboratoire AIM, CEA, Universit\'{e} Paris Diderot, IRFU/Service d'Astrophysique, Bat. 709, 91191 Gif-sur-Yvette, France \\
}
}

\maketitle{}

 
\begin{abstract}

We present a line survey of the ultra-luminous infrared galaxy Arp 220, taken with 
the newly installed SEPIA Band 5 instrument on APEX. We illustrate the capacity of 
SEPIA to detect the 183.3 GHz H$_2$O 3$_{1,3}$-2$_{2,0}$ line against the 
atmospheric H$_2$O absorption feature. We confirm the previous detection of the HCN(2$-$1) line, 
and detect new transitions of standard dense gas tracers such as HNC(2$-$1), 
HCO$^+$(2$-$1), CS(4$-$3), C$^{34}$S(4$-$3), HC$_3$N(20-19). We also detect HCN(2$-$1) v2=1 
and the 193.5 GHz methanol (4$-$3) group for the first time. The absence of time variations in the 
megamaser water line compared to previous observations seems to rule out an AGN nuclear origin for the line.
It could, on the contrary, favor a thermal origin instead, but also possibly be a sign that the 
megamaser emission is associated with star-forming cores washed-out in the beam. 
We finally discuss how the new transitions of HCN, HNC, HCO$^+$
refine our knowledge of the ISM physical conditions in Arp~220. 


\end{abstract}
  


\vspace{50pt}
\section{Introduction}

Arp 220 is one of the nearest (z=0.018) ultra-luminous infrared galaxy (L$_{\rm IR}$ $>$ 10$^{12}$ \lsun). Many studies
have analysed its massive molecular component and the physical conditions powering its intense infrared (IR) luminosity. 
Diffuse (e.g., low-J CO) and dense (HCN, HNC, HCO$^+$) gas tracers have been detected 
\citep[][hereafter K08, I10 and G09]{Krips2008,Imanishi2010,Greve2009}, providing us with estimates of the molecular 
gas masses and densities. Physical conditions of the interstellar medium (ISM) can also be probed 
using maser (Microwave Amplified Stimulated Emission Radiation) emission \citep{Lo2005}.
Even if the 22 GHz water megamaser emission has been detected in more than 150 extragalactic 
objects so far \citep[it usually arises from Active galactic nucleus (AGN) central engines in Seyfert 2 or
LINER galaxies; e.g.][]{Braatz1996,Lo2005}, this line was not detected towards Arp 220.
Instead, Arp 220 exhibits OH megamaser emission \citep{Rovilos2003}. 
Using the IRAM 30-m telescope, \citet{Cernicharo2006} (hereafter C06) have detected the H$_2$O(3$_{1,3}$-2$_{2,0}$) line 
at 183.3 GHz. The line, originating from a transition at a low energy level above ground state, is of sufficient 
luminosity to be classified as a megamaser. Given the absence of the 22 GHz water emission, 
they suggest that the 183 GHz emission is arising from the relatively spatially extended starburst (from about 10$^6$ 
star-forming cores) as opposed to denser molecular gas in the AGN nuclei.

The SEPIA (Swedish-European Southern Observatory PI receiver for APEX) Band 5 (159$-$211 GHz) instrument is a dual polarization 
two-sideband receiver based on the ALMA Band 5 receivers \citep{Billade2012}. The Science Verification (SV) 
campaign was carried out on the Atacama Pathfinder Experiment (APEX) antenna in 2015. The galaxy Arp~220 
was observed as part of the SV campaign because of its importance as a local benchmark for high-redshift galaxies. 
The goals of the following analysis are to {\it i)} compare the H$_2$O(3$_{1,3}$-2$_{2,0}$) and 
HCN(2$-$1) lines observed with SEPIA Band 5 with previous observations in Arp~220 in order to test the 
calibration of the instrument and monitor the water line emission, {\it ii)} target millimeter wavelength 
CH$_3$OH maser emission and analyse new transitions of standard dense gas tracers such as HNC, CS, HC$_3$N or 
CH$_{\rm 3}$CN now accessible with SEPIA Band 5. We describe the observations and data reduction technique 
in $\S$2 and present a detailed overview of the various lines detected in $\S$3. We analyse the observations in $\S$4.

\begin{table*}
\caption{Characteristics of the various fitted lines (Gaussian fitting).}
\label{Line_characteristics}
 \centering
 \begin{tabular}{ccccccccccc}
\hline
\hline
\vspace{-5pt}
&\\
Setup 				&Line								&	$\nu$$_{rest}$ $^{a}$ & Vel. Offset $^{b}$ & FWHM 			& T$_{\rm peak}$ $^{c}$   & {\it $\int$ Sdv} 	    &	$L'$				&  \multicolumn{2}{c}{Previous measurements $^{(d)}$}\\
	  				& 									& 	(GHz) 		          &(km~s$^{-1}$) 	& (km~s$^{-1}$)	& (mK)  		     &	(Jy km s$^{-1}$)   &	(10$^7$ K~km~s$^{-1}$ pc$^2$)	& FWHM & $L'$ \\
\vspace{-5pt}
&\\
\hline
\vspace{-5pt}
&\\
Tuning 1 $^{e}$			&	HCN(2$-$1)  						&  177.26  			& +99 $\pm$ 13	&  452 $\pm$ 25 		  &      9.6 $\pm$ 0.4  	& 157 $\pm$ 11 	& 80 $\pm$ 6 	& 530 $\pm$ 20 & 88 $\pm$ 26 \\
					&	HCO$^+$(2$-$1)   					&  178.37  			& 0 $^{g}$ 			&  ~~484 $\pm$ 25 $^{f}$    &      4.6 $\pm$ 0.3		& 79 $\pm$ 7		& 40 $\pm$ 3 	& - & -  \\

\vspace{-7pt}
&\\

Tuning 2 $^{f}$			&	HNC(2$-$1)  		    				& 181.32 				& +80 $\pm$ 11	&  484 $\pm$ 25 		  & 	  6.6 $\pm$ 0.3		& 116 $\pm$ 8 	& 57 $\pm$ 4 & - & -  	\\
					&	HC$_{\rm 3}$N(20$-$19)				& 181.94				& +84 $\pm$ 29	&  262 $\pm$ 54  		  &      1.9 $\pm$ 0.2 		& 18 $\pm$ 4 	&  9 $\pm$ 2  & - & - 	\\
					&	H$_{\rm 2}$O(3$_{1,3}$$-$2$_{2,0}$)	& 183.31       			& +67 $\pm$ 14	&  332 $\pm$ 31      &	  4.1 $\pm$ 0.3		& 50 $\pm$ 6 	& 24 $\pm$ 3 & 310 & 25 	\\
					&	C$^{34}$S(4$-$3)					& 192.82  				&+128 $\pm$ 42	&  495 $\pm$ 92		  &      1.3 $\pm$ 0.1 		& 23 $\pm$ 5    & 10 $\pm$ 2 & - & -  	\\
					&	CH$_3$OH(4$-$3)					& 193.45  				& +25 $\pm$ 27	&  333 $\pm$ 42		  &      2.0 $\pm$ 0.2		& 23 $\pm$ 5    & 10 $\pm$ 2 & - & -  	\\
					&	CS(4$-$3)				 			& 195.95  				& +60 $\pm$ 12	&  418 $\pm$ 25 		  &      4.0 $\pm$ 0.2   	& 60 $\pm$ 5    & 25 $\pm$ 2 &	 - & -     \\
\vspace{-5pt}

&\\
\hline
\end{tabular}
\begin{list}{}{}
\item[{\bf Notes --}] 
$^{(a)}$ z=0.018
$^{(b)}$ Velocity Offset of the fitted peak with the systemic velocity
$^{(c)}$ in T$^*_A$ scale
$^{(d)}$ Same units. References: K08 and C06
$^{(e)}$ Frequency ranges observed = [172.2-176.2 GHz] (LSB) and [184.2-188.2 GHz] (USB)
$^{(f)}$ Frequency ranges observed = [177.1-181.1 GHz] (LSB) and [189.1-193.1 GHz] (USB)
$^{(g)}$ The HCO$^+$(2$-$1) line is blended with the HCN(2$-$1) v$_2$=1 line. The central velocity and width of the HCO$^+$ line are fixed for the fitting.

\end{list}
 \end{table*} 

\section{Observations and data reduction}

SEPIA Band 5 covers the frequency range 159$-$211 GHz. 
The lower and upper sideband (LSB and USB) are separated by 12 GHz. Each sideband 
is recorded by 2 XFFTS (eXtended bandwidth Fast Fourier Transform Spectrometer) 
units of 2.5 GHz each, with a 1 GHz overlap. 
The spectral resolution of the data can reach up to 0.065 km~s$^{-1}$ and the beam 
size is $\sim$ 35\arcsec\ (equivalent to 12.7 kpc for Arp~220). 
We use a Jy/K factor of 34 to convert between antenna temperature T$^*_A$ and 
flux density\footnote{http://www.eso.org/sci/activities/apexsv/sepia/sepia-band-5.html}. We 
refer to \citet{Billade2012} and Immer et al (in prep) for more details on the receiver. 
The performance of the instrument will be presented in a future paper.
We conducted observations towards Arp~220 (Project ID: 095.F-9801; PI: Galametz) 
using the wobbler switch mode in 2015 July and September with a precipitable water 
vapour (pwv) $<$ 0.7 mm. The observations were carried out for a total time of 6.2 hours, with 2.8 hours on source. 
Pointings were regularly checked on Carina and a calibration scan was taken every 10 min.
Data were reduced using CLASS/Gildas\footnote{Gildas Continuum and Line Analysis Single-dish 
Software, http://www.iram.fr/IRAMFR/GILDAS}. We removed a linear baseline from each individual spectrum before averaging the data.
The rms reached for the water and HCN observations are 0.7 mK and 1.2 mK respectively, at a 50 km~s$^{-1}$ resolution.

\begin{figure}
    \centering
    \vspace{-15pt}
    \begin{tabular}{c}
\vspace{-2pt}
\hspace{-5pt} \includegraphics[width=8cm]{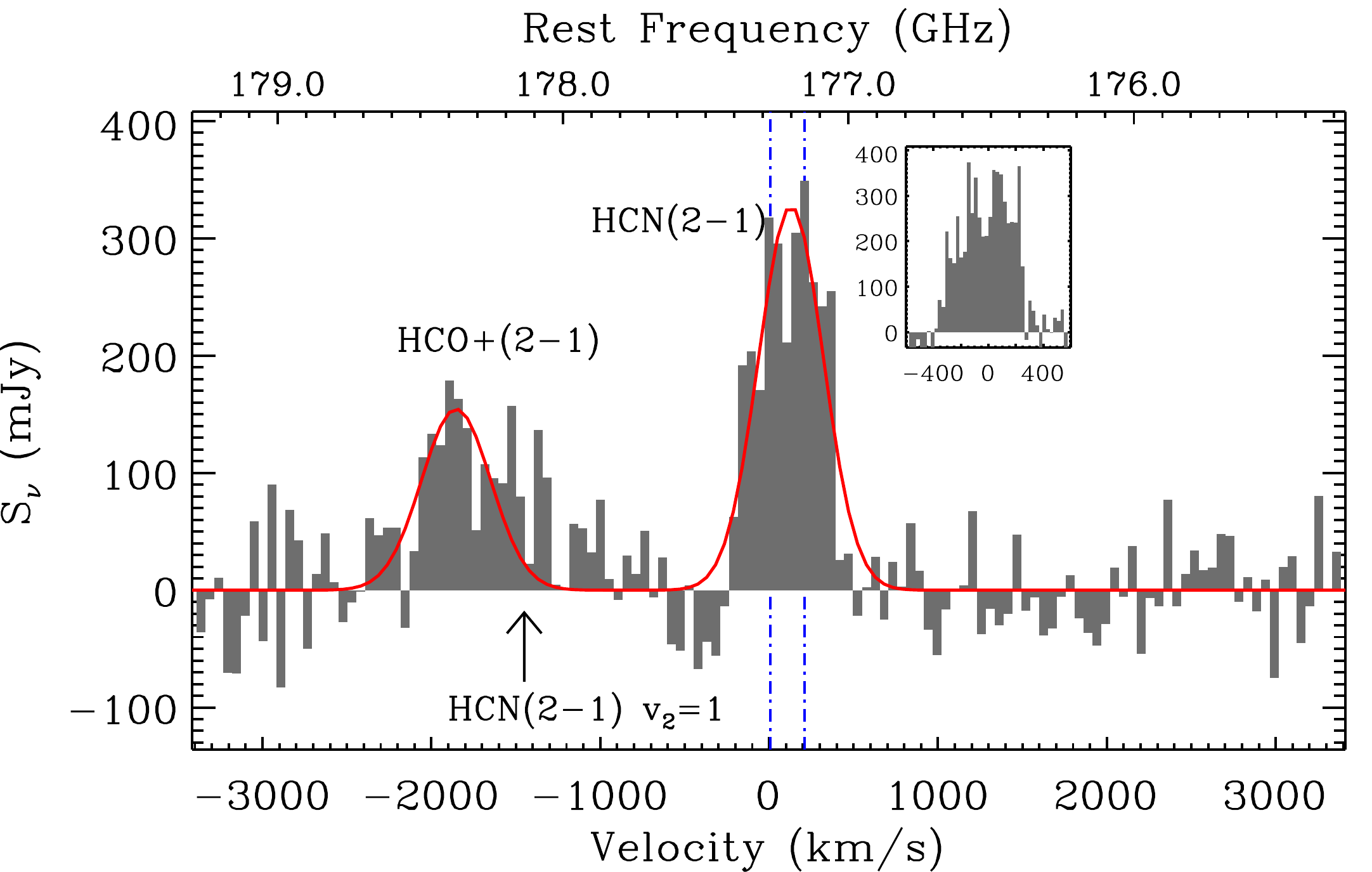} \\
\vspace{-2pt}
\hspace{-5pt} \includegraphics[width=8cm]{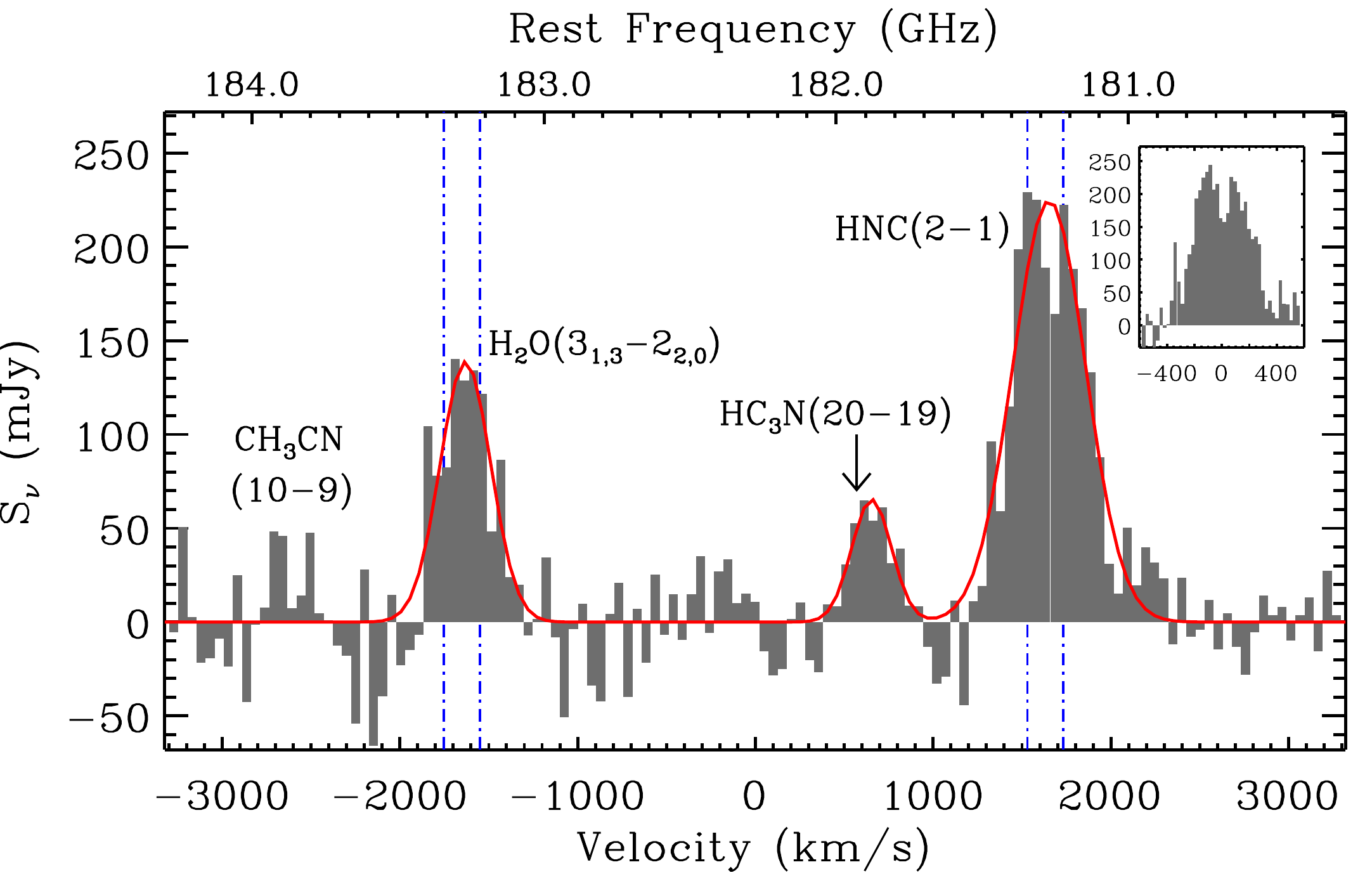}  \\
\vspace{-2pt}
\hspace{-5pt} \includegraphics[width=8cm]{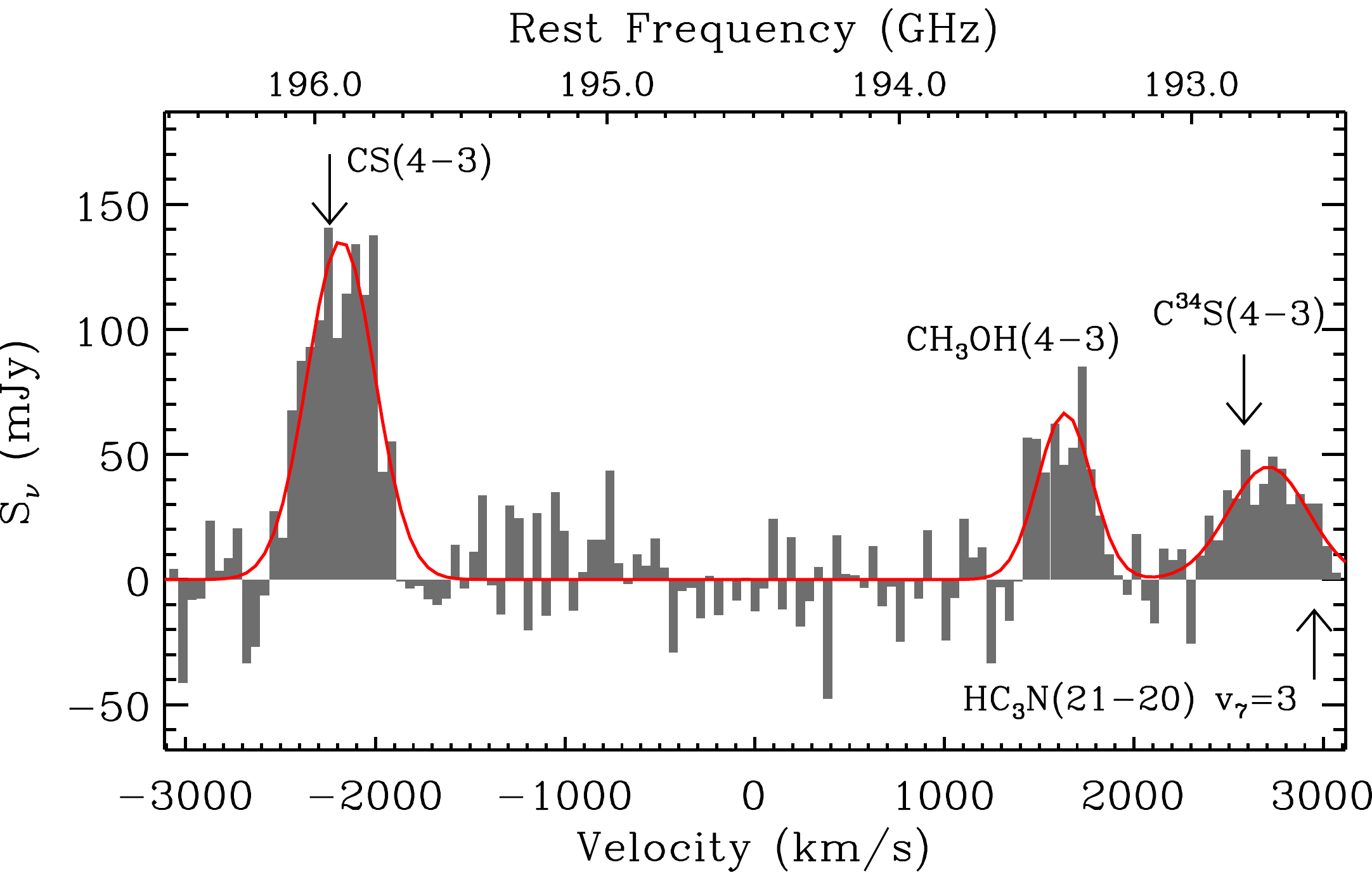}  \\
         \end{tabular}
    \caption{SEPIA Band 5 observations toward Arp~220 (6.2 hours). The plotted spectral resolution
    is 50 km~s$^{-1}$. We observe strong water and methanol emission as well as many tracers of 
    the dense gas (HCN, HNC, HCO$^+$, CS, HC$_{\rm 3}$N, C$^{34}$S). 
    Insets show the HCN and HNC double-peak profiles at a 25 km~s$^{-1}$ resolution 
    (x-axis,: velocity relative to the center of the line). We overlay the position of their respective 
    peaks on the H$_2$O, HCN and HNC lines (blue). }
    \label{Spectra}
\end{figure}

\vspace{-5pt}
\section{Overview of the detected lines}

In this section, we present the lines detected by SEPIA Band 5 receiver toward Arp~220. 
We show the spectra in Fig.~\ref{Spectra} and describe the lines in order of decreasing peak temperature. 
We provide the parameters of the Gaussian fits of the lines in Table~\ref{Line_characteristics}. Note, however, that 
some of the lines have broad non-Gaussian profiles.\\

\noindent {\it HCN and HNC - }The hydrogen cyanide and isocyanide are very sensitive 
tracers of the dense molecular phase. The HCN luminosity is in particular used as a robust 
proxy for the star formation rate \citep[e.g.][]{Privon2015}. Both HCN and HNC have been 
widely observed in Arp~220. HCN(2$-$1) is the strongest line in our observations. Its line width (452 km~s$^{-1}$) 
is consistent with that of the HCN(3$-$2) line estimated by C06 but smaller than the linewidth of the HCN(2$-$1) already observed
by K08. The two HCN(2$-$1) flux densities, however, match within the uncertainties (Table~\ref{Line_characteristics}). 
The right side of the HCN profile looks truncated. The width of the HNC(2$-$1) line is more consistent with the HCN linewidths estimated in K08.
Its peak temperature is $\sim$1.5 times lower than that of HCN(2$-$1) (see $\S$4.3).
Both HCN and HNC exhibit a double-peak profile. The HCN(2$-$1) line has a stronger peak at higher velocities 
(+170 km~s$^{-1}$ shift compared to the systemic velocity). This asymmetry has already been 
found for HCN in the 3$-$2 and 4$-$3 transitions while the 1$-$0 line shows two symmetric peaks.  
On the contrary, the HNC(2$-$1) line shows a stronger peak at lower velocities ($-$50 km~s$^{-1}$ shift compared to the systemic velocity)
already observed in the 1$-$0 transition. Arp~220 possesses two nuclei, each of them being separated 
by about 1\arcsec\ and resolved in the radio continuum at 1 and 6mm by \citet{Sakamoto1999} and \citet{Rodriguez-Rico2005}. 
Our observations could indicate that a larger fraction of the HCN emission is arising from the red-shifted eastern nucleus 
(as already suggested by I10) and that a larger fraction of the HNC emission is arising from the 
blue-shifted western nucleus. However, recent observations of HCN with ALMA have shown that a double-peak 
velocity profile is observed towards each individual nucleus \citep{Scoville2015,Martin2016}. The double-peak profiles
are thus probably linked with self- and continuum absorption along the line of sight. 
Follow-up imaging with ALMA would help disentangle the two effects in the future, when Band 5 becomes available on ALMA.

\vspace{3pt}
\noindent {\it HCO$^+$ - }The formylium line traces gas $\sim$100-500 times denser than the low-J CO. 
\citet{GraciaCarpio2006} suggested HCO$^+$ as a more robust dense gas tracer
in (U)LIRGs than HCN because of the enhanced HCN-to-CO ratio of these galaxies linked with
IR pumping or chemical enhancement. Previous observations of HCO$^+$ in Arp~220 in the 1$-$0, 3$-$2 and 4$-$3 transitions 
are reported in \citet{GraciaCarpio2006}, G09, I10 and \citet{Aladro2015}. 
We detect HCO$^+$(2$-$1) ($\nu$$_{\rm rest}$ = 178.4 GHz) for the first time toward Arp~220 (Fig.~\ref{Spectra}, top). 
The HCO$^+$ line is blended with the HCN(2$-$1) line in its v$_2$=1 vibrational state ($\nu$$_{\rm rest}$ = 178.1 GHz; 
HCN$_{vib}$-to-HCN(2$-$1) line ratio $<$ 0.5). In order to properly fit the HCO$^+$ profile, we thus use a Gaussian 
positioned at the expected frequency of the line and assume a similar width as the well constrained HNC line (i.e. 484 km~s$^{-1}$). 
We find a peak temperature of 4.6 mK (0.16 Jy), twice lower than that of HCN.

\vspace{3pt}
\noindent {\it CS - } Up to the 4$-$3 transition, the carbon monosulfide molecule traces gas densities of 
10$^5$-10$^6$ cm$^{-3}$ while higher transitions trace 
densities $>$ 10$^6$ cm$^{-3}$ \citep{Bayet2009,Shirley2015}. In contrast with HCN or HNC,
CS seems to be less affected by radiative excitation from X-ray dominated regions (XDRs) 
or by IR pumping \citep[][]{Meijerink2007}.
Previous observations of CS(2$-$1), (3$-$2), (5$-$4) and (7$-$6) toward Arp~220 are reported in G09 and \citet{Aladro2015}.
For the first time, the CS(4$-$3) line ($\nu$$_{\rm rest}$ = 195.9 GHz) is prominently detected in Arp~220 (Fig.~\ref{Spectra}, bottom). 
Its width is three times larger than the average line width of CS(4$-$3) lines observed in local (D $<$ 10Mpc) 
galaxies \citep{Bayet2009}. The CS(4$-$3) profile resembles the asymmetric broad profiles 
already observed in other transitions, with a redshifted stronger component (G09). 
The asymmetry could, as for HCN, be the sign of self-absorption:
\citet{Scoville2015} also detected a double-peak profile in the CS line in both nuclei with ALMA. 
\citet{Zhang2014} have shown that a strong correlation (that extends over 
eight orders of magnitude) exists between the CS luminosity and the IR luminosity, 
making the CS luminosity an excellent proxy of the star formation rate. 
In Arp~220, the L'$_{CS(7-6)}$-to-L'$_{CS(4-3)}$ ratio is equal to 0.76. Using the calibration of \citet{Zhang2014}, 
this leads to a L$_{\rm IR}$ estimate of 1.6$\times$10$^{\rm 12}$ \lsun, consistent with the L$_{\rm IR}$ determined by 
\citet{Rangwala2011} using \hersc/SPIRE data (i.e. 1.8$\times$10$^{\rm 12}$ \lsun).
The relation linking L$_{CS(4-3)}$ and L$_{\rm IR}$ can now be calibrated with SEPIA Band 5.

\vspace{3pt}
\noindent {\it H$_2$O - } Submillimeter water line emission is considered to be 
an important gas cooling mechanisms in warm molecular clouds and the H$_{\rm 2}$O line ($\nu$$_{\rm rest}$ = 183.3 GHz)
 only requires moderate densities (n(H$_{\rm 2}$) = 10$^4$$-$10$^5$ cm$^{-3}$) and low temperatures (T$\sim$40K) to be excited \citep{Cernicharo1994}.
We confirm the detection of H$_{\rm 2}$O(3$_{3,1}$$-$2$_{2,0}$) emission in Arp~220 (Fig.~\ref{Spectra}; middle),
with a peak flux of 0.14 Jy and a width at half-power 
of 332$\pm$31 km~s$^{-1}$. These values are consistent with the peak flux (0.17 Jy) and 
width at half-power (310 km~s$^{-1}$) obtained with IRAM data by C06 (see section 4.1 for further analysis).

\vspace{3pt}
\noindent{\it CH$_{\rm 3}$OH -} Extragalactic methanol maser emission has been detected toward 
many objects \citep{Henkel1987,Sjouwerman2010}. In particular, millimeter methanol transitions are known to be sensitive (but sometimes degenerate) 
tracers of temperature and density. Millimeter CH$_{\rm 3}$OH emission has been detected in Arp~220 by \citet{Martin2011}, particularly 
the 5$-$4 group (241.8 GHz).  \citet{Salter2008} also report an absorption feature from the 6.7 GHz methanol line.
Our SEPIA observations give us an access to the 193.5 GHz methanol (4$-$3) group (Fig.~\ref{Spectra}, bottom). The peak flux is 68 mJy. The water
and methanol lines have similar widths (equal to 333 km~s$^{-1}$ for CH$_{\rm 3}$OH). We note that this 
is at the lower end of the line width range (280 to 590 km~s$^{-1}$) estimated from thermal molecules in Arp~220 
by G09. 

\vspace{3pt}
\noindent {\it HC$_{\rm 3}$N -} Molecules from the cyanopolyyne family like HC$_{\rm 3}$N are standard 
tracers of the dense gas in Galactic star forming regions {but are also detected in 
extragalactic environments \citep{Aladro2011,Lindberg2011} in many transitions} (from 9$-$8 to 28$-$27). 
Arp~220 is one of the very few ``HC$_{\rm 3}$N-luminous'' galaxies \citep[i.e. that the HC$_{\rm 3}$N(10$-$9) line 
intensity is at least 15$\%$ of the HCN(1$-$0) line intensity;][]{Aalto2002,Martin2011}.
The HC$_{\rm 3}$N(20$-$19) line is narrow and detected at 16\% of the 
HNC(2$-$1) luminosity. Its profile looks symmetric but its peak is shifted. This is 
consistent with the assymetry observed in the 10$-$9 transition by \citet[][]{Aalto2002} and 
could suggest that most of the emission emerges from the eastern nucleus.

\vspace{3pt}
\noindent {\it C$^{34}$S -} We observe a strong and broad C$^{34}$S(4$-$3) line at 192.8 GHz. 
The line is unfortunately at the edge of the velocity range observed.
A HC$_3$N vibrationally excited line resides in a similar frequency range (192.6 GHz; see Fig.~\ref{Spectra}, bottom)). Vibrationally 
excited emission of this species was already reported in \citet[][]{Martin2011}. The line should not contribute 
significantly due to its high energy level. 

\vspace{3pt}
\noindent {\it CH$_{\rm 3}$CN }- In Arp~220, the acetonitrile line was detected in both the ground and 
vibrationally excited states by \citet{Martin2011}. Our observations cover the CH$_{\rm 3}$CN(10$-$9) transition, 
for which we do not have a clear detection (2-$\sigma$; Fig.~\ref{Spectra}, middle). 
It is then difficult to say if the line is double-peaked as the HCN line.

\section{Analysis}

This section presents a first analysis of the data focused on the water, methanol, HCN, HNC and HCO$^{+}$ lines.
A more detailed analysis of the present and complementary datasets will be presented in a following paper.

\subsection{Nature of the 183 GHz water emission} 

The 183 GHz H$_{\rm 2}$O(3$_{3,1}$$-$2$_{2,0}$) line was observed by C06. Its 
intense luminosity suggested a megamaser (i.e. $>$ 10$^6$ times the 
luminosity of Galactic analogues) origin. We re-observe the H$_{\rm 2}$O line to monitor
its possible time variation if it is excited by AGN activity. 
We derive the isotropic line luminosity using:

\begin{equation}
\frac{L}{[L_{\rm \odot}]} = 10^{-3}\frac{1}{1+z}\frac{\nu_{\rm rest}}{\rm [GHz]}\frac{\rm D^{2}}{\rm [Mpc^2]}\frac{\int{Sd v}}{\rm [Jy~km~s^{-1}]}
\end{equation}

\vspace{5pt}
\noindent with z being the redshift of the source, $\nu$$_{\rm obs}$ the rest frequency of the line, D the distance to the source 
and $\int${\it S d}v the integrated flux density. We find an isotropic luminosity of 4.6 $\times$ 10$^{4}$ \lsun. This is 
million times more luminous than the typical 10$^{-3}$ \lsun\ Galactic 22 GHz maser emission \citep{Genzel1977}. In 
Galactic targets, maser emission is typically stronger at 22 GHz compared to 183 GHz. Our 
detection is thus also more than a million times more luminous than the typical Galactic 183 GHz maser emission. 
It is also 2-3 orders of magnitude higher than 22 GHz masers detected in active galaxies \citep[][among others]{Braatz1996}.  
The Seyfert 2 galaxy NGC 3079 was the first extragalactic object where 183 GHz water emission was observed \citep{Humphreys2005}. 
The line was narrow ($<$ 50 km~s$^{-1}$), and interpreted as arising from a portion of the AGN central engine. In contrast, the 183 GHz water
line in Arp~220 is much broader than Galactic 183 GHz analogues \citep{Waters1980} or than in NGC 3079. It has a
profile similar to the HCN or HNC lines (the FWHM is 1.4 times narrower for H$_2$O) that could suggest a thermal origin. 
The single-dish observations do not allow to disentangle between an extended/thermal and a nuclear/maser origin. 
AGN water megamasers at 22 GHz typically vary by factors of a few. Yet, we do not find 
variation between the 183 GHz water emission measured in 2005 by C06 and with SEPIA in 2015: the line has a similar width and 
luminosity. This lack of variability probably argues against the emission originating from 
an AGN nuclei in Arp~220 and provides weight to the proposal by C06 that the water megamaser
emission in Arp~220 is associated with star formation and arises from individual cores (estimated at 10$^6$ by C06). 
The variabilities in these individual cores could then be washed-out in a single-dish spectrum. 
OH megamasers are, in the same way, not commonly reported to be highly variable in single-dish 
studies \citep[variability $\le$ 30\%; e.g.][]{Darling2002}. More SEPIA observations are needed to monitor possible variations in the water line.
We finally note that \citet{Gonzalez-Alfonso2012} suggested that the extreme H$_2$O abundances in Arp~220 
could also be the result of a hot core chemistry \citep[i.e. characterized by the evaporation of icy grain 
mantles; see][among others]{Charnley1995}.

\subsection{Nature of the 193.5 GHz methanol emission}

Extragalactic methanol maser emission was only classified as megamaser in NGC~1068 \citep[84.5 GHz;][]{Wang2014} 
and Arp~220 \citep[36.2 and 37.7 GHz;][]{Chen2015}. We find an isotropic line luminosity of 2.3$\pm$0.3 $\times$ 10$^{4}$ \lsun\
for the 193.5 GHz methanol line, here again suggesting a megamaser origin. The luminosity is 9 orders of magnitude 
higher than typical Galactic or Large Magellanic Cloud methanol masers \citep[][]{Breen2010,Ellingsen2010}. The line is much 
broader than typical extragalactic methanol masers or than the megamaser emission detected in Arp~220 by \citet{Chen2015}, 
which could also favor a thermal origin. We note that the luminosity of the CH$_3$OH(2$_k$$-$1$_k$) line 
\citep[7.5 $\times$ 10$^7$ K~km~s$^{-1}$~pc$^2$;][]{Aladro2015} is lower than the luminosity of the 4$-$3 transition. In LTE conditions, 
the CH$_3$OH(4$-$3) is expected to be more difficult to excite than the lower transitions if it was resulting from a pure thermal excitation. 
We cannot, however, rule out a thermal origin in the case of highly non-LTE excitation conditions.
Getting more CH$_3$OH lines at other frequencies would be necessary to understand the origin of the emission.

\subsection{Line ratios and excitation conditions} 

\begin{figure}
    \centering
        \vspace{-15pt}
    \begin{tabular}{cc}
\hspace{-10pt} \includegraphics[height=4.7cm]{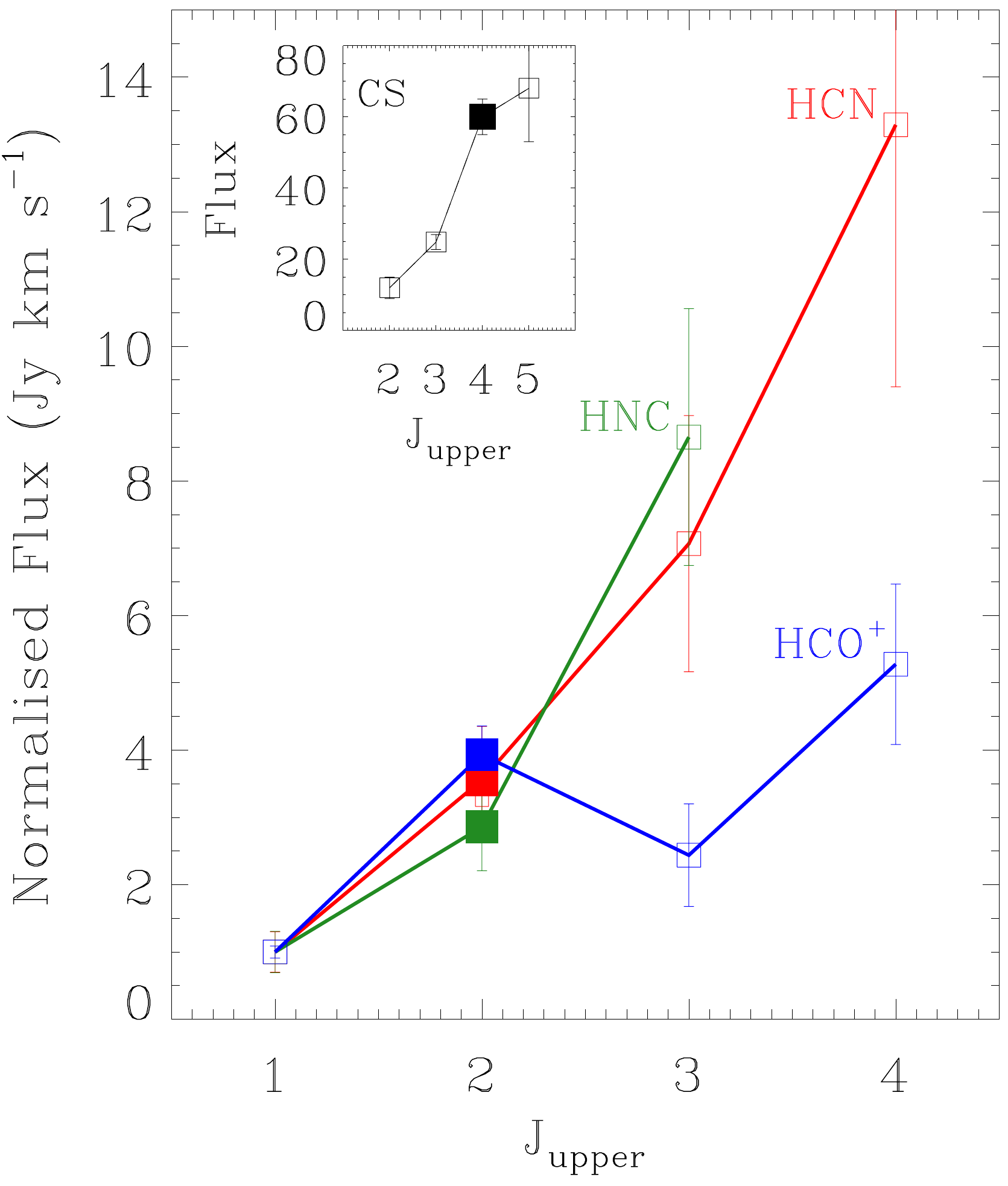}  &
\hspace{-12pt} \includegraphics[height=4.7cm]{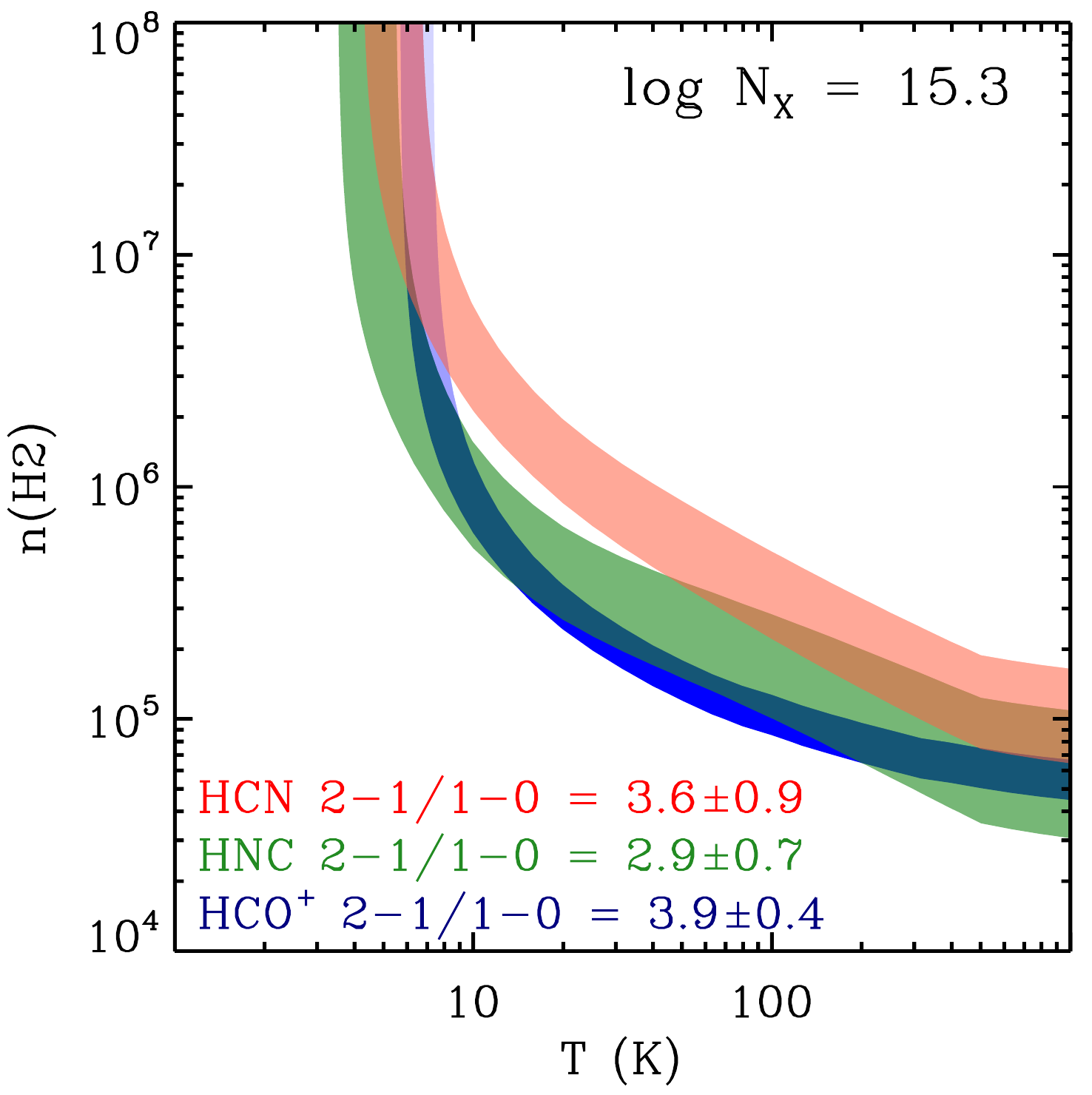}  \\
         \end{tabular}
                 \vspace{-10pt}
    \caption{{\it Left:} Rotational ladders of HCN, HNC and HCO$^{+}$ (fluxes are normalised to that of the 1$-$0 transition). 
    Inset: CS rotational ladder (not normalised). Filled squares are from this work.
    {\it Right:} RADEX modeling of the HCN, HNC and HCO$^+$ 2$-$1-to-1$-$0 ratios (units of Jy km~s$^{-1}$) 
    in a kinetic temperature (in K) versus H$_2$ gas density (in cm$^{-3}$) diagram. The bands indicate the parameter 
    space covered by each ratio, including uncertainties. For the model presented here, the column 
    density of each molecular species (N$_X$) is fixed to the N(HCN) estimated by \citet{Rangwala2011}. 
}
    \label{Radex}
\end{figure}

Figure~\ref{Radex} shows the HCN, HNC, HCO$^+$ and CS spectral line energy distributions (SLEDs).
Fluxes are gathered from this work, \citet{Aalto2009} and G09 (we use the rms weighted 
averages of the reported line luminosities). Arp~220 is one of the few objects where the 
HNC/HCN(1$-$0) and (3$-$2) line ratios are close or above unity \citep[][G09]{Huettemeister1995}. 
In contrast, we find a HNC/HCN(2$-$1) luminosity ratio well below unity ($\sim$0.7) that suggests 
a more standard HNC chemistry. A proper HNC(2$-$1) survey would help assess how the HNC/HCN(2$-$1) 
ratio in Arp~220 compares with other nearby galaxies. 
We use the non-LTE radiative transfer code Radex \citep{vanderTak2007} to generate diagnostic plots 
of the HCN, HNC and HCO$^+$ 2$-$1/1$-$0 ratios for various kinetic temperatures and H$_2$ number 
densities. We simply assume here that the CMB 2.73 K emission dominates the background emission 
in the mm regime. Figure~\ref{Radex} shows the T-n(H$_2$) parameter spaces covered by the 2$-1$/1$-$0 ratios if
we assume the column density derived by \citet[][]{Rangwala2011} from HCN high-J absorption lines, i.e. 
2$\times$10$^{15}$ cm$^{-2}$. We find that the domains covered by the 3 species barely overlap. 
Increasing the background blackbody temperature shifts all solutions to lower T and n(H$_2$) but do not make the domains converge.
A solution domain is reached when we assume higher column densities (10$^{16}$ cm$^{-2}$). It requires lower 
n(H$_2$) ($\sim$10$^5$ cm$^{-3}$) than those derived in \citet[][]{Rangwala2011}, but similarly high 
temperatures ($>$ 300 K). Another solution domain is reached for n(H$_2$)$\sim$10$^7$ cm$^{-3}$
but the temperatures ($<$10K) requested are too low to be realistic (much lower than the temperature of the dust in Arp~220).
G09 and I10 have showed that using HCO$^+$ line ratios results in n(H$_2$) estimates 1-2 orders of magnitude lower than those derived 
from HCN. In contrast, our moderate HCN/HCO$^+$(2$-$1) line ratio does not
lead to such a strong discrepancy. We note however that the estimated HCO$^+$(2$-$1) 
flux is high (Figure~\ref{Radex}, left). Because the HCN(2$-$1) v$_2$=1 line partly 
contributes to the flux, the HCO$^+$(2$-$1) flux estimate is uncertain and should 
probably be considered as an upper limit. 

The high HCN(2$-$1/1$-$0) renders the ratios difficult to reconcile with our simple model. 
In fact, many additional physical processes are at work in Arp~220. First, its HCN emission
is thought to be enhanced via IR pumping. \citet{Rangwala2011} suggested, for instance, 
that the HCN high-J absorption lines might be populated via an IR pumping of the vibrational 
states. Vibrationally excited HCN is indeed observed toward Arp~220 \citep{Aalto2015,Martin2016} 
and our detection of the HCN(2$-$1) v$_2$=1 line is an additional sign that IR pumping processes 
are at work in Arp~220. These processes should also affect the HNC line, as suggested by \citet{Aalto2007}.
Second, our simple model neglects the effects of a hot dust background that could be non 
negligible (even on the line ratios) in extreme environments such as Arp~220 \citep{Tunnard2016}.
Finally, the double-peak profiles of the HCN and HNC lines are strong signs of self- and continuum absorption
whose effects on the various transitions of the lines still need to be properly quantified \citep{Papadopoulos2010,Rangwala2015,Martin2016}.
These processes will be investigated and integrated in a more complex modelling in a following paper.

\vspace{-5pt}
\section{Conclusions}

The paper presents new observations toward Arp~220 using the SEPIA Band 5
instrument on APEX. We demonstrate the capacity of SEPIA to observe the 183 GHz water and 193.5 GHz 
methanol lines in this extreme extragalactic environment. Their luminosities classify them as megamaser but
the absence of variation in the water line could rule out an AGN nuclei origin.
SEPIA is an ideal instrument to monitor variations of maser lines.
Our observations complement the low-J SLEDs of standard molecular species, with lines such as HCN(2$-$1), 
HNC(2$-$1) and HCO$^+$(2$-$1) or CS(4$-$3) lines. Arp~220 exhibits a more standard HNC/HCN(2$-$1) ($<$1) 
and moderate HCN/HCO+(2$-$1) line ratio compared to other transitions. 
The HCN and HNC lines exhibit double-peak profiles probably linked 
with self- and continuum absorption. New transitions of HC$_3$N and CH$_3$CN and a vibrational excited HCN line are 
also detected with the instrument.

\vspace{-10pt}
\section*{Acknowledgments}
We would like to thank the referee for his/her very useful comments and suggestions 
and both the SEPIA team and the APEX operating team onsite for their hard work in making 
the SEPIA commissioning a success. 
APEX is a collaboration between the Max-Planck-Institut fur Radioastronomie, the European 
Southern Observatory (ESO), and the Onsala Space Observatory. SEPIA is a collaboration between Sweden and ESO. 
Z.Y.Z. acknowledges support from the European Research Council (Advanced Grant COSMICISM).

\vspace{-10pt}
\bibliographystyle{mn2e}
\bibliography{/Users/maudgalametz/Documents/Work/Papers/mybiblio.bib}

\end{document}